%% file: ms.tex
\pdfoutput=1

\documentclass{llncs}
\usepackage[utf8]{inputenc}
\usepackage[T1]{fontenc}
\usepackage{microtype}
\usepackage{times}

\usepackage{amsmath}
\usepackage{amssymb}
\usepackage{balance}
\usepackage{dsfont}
\usepackage{stmaryrd}
\usepackage{enumerate}
\usepackage{caption}
\usepackage{subcaption}
\usepackage{capt-of}
\usepackage{pgf}
\usepackage{tikz}
\usepackage{algorithm,algorithmic}

\usepackage{booktabs}
\usepackage{graphicx}
\usepackage{epstopdf}

\usetikzlibrary{automata}
\usetikzlibrary{arrows,shapes,calc}
\usetikzlibrary{positioning}
\usepackage{array}
\usepackage{booktabs}
\usepackage{wrapfig}

\input{macrosLNCS}

\newcommand{\idol}{\textsf{IDOL}}

\newcommand{\syn}[1]{#1}

\renewcommand{\vec}[1]{\mathbf{#1}}

\newcommand{\frc}{\mathfrak{c}}

\newcommand{\fnt}{\sim_{\h,\varepsilon}^{F}}
\newcommand{\bnt}{\sim_{\h,\varepsilon}^{B}}
\newcommand{\ft}{\sim_{\h,\varepsilon}^{F^\ast}}
\newcommand{\bt}{\sim_{\h,\varepsilon}^{B^\ast}}

\newcommand{\efnt}{\sim_{\h}^{F}}
\newcommand{\ebnt}{\sim_{\h}^{B}}
\newcommand{\eft}{\sim_{\h}^{F^\ast}}
\newcommand{\ebt}{\sim_{\h}^{B^\ast}}

\newcommand{\wpf}{\wp^H_{i,j}}
\newcommand{\wpb}{\wp^\h_{i,j}}

\newcommand{\g}{\mathcal{G}}

\newcommand{\dvars}{\mathcal{S}}

\newcommand{\nf}{\mathcal{N}}

\newcommand{\argmin}{\operatornamewithlimits{argmin}}

\newcommand{\hs}{\hat{\sigma}}
\newcommand{\hso}{\hat{\sigma}_\ast}
\newcommand{\calp}{\mathcal{P}}

\newcolumntype{H}{>{\setbox0=\hbox\bgroup}c<{\egroup}@{}}

\begin{document}

\title{Guaranteed Error Bounds on Approximate Model Abstractions through Reachability Analysis}

\institute{
	Microsoft Research
	\&
	University of Oxford, UK
    \and
	IMT School for Advanced Studies Lucca, Italy
	\and
	DTU Compute, Denmark
}

\author{
	Luca Cardelli\inst{1}\and
	Mirco Tribastone \inst{2} \and
	Max Tschaikowski \inst{2} \and
	Andrea Vandin \inst{3}
}
\authorrunning{Luca Cardelli, M. Tribastone, M. Tschaikowski, and A. Vandin}

\maketitle

\begin{abstract}
It is well known that exact notions of model abstraction and reduction for dynamical systems may not be robust enough in practice because they are highly sensitive to the specific choice of parameters. In this paper we consider this problem for nonlinear ordinary differential equations (ODEs) with polynomial derivatives. We introduce approximate differential equivalence as a more permissive variant of a recently developed exact counterpart, allowing ODE variables to be related even when they are governed by nearby derivatives. We develop algorithms to (i) compute the largest approximate differential equivalence; (ii) construct an approximate quotient model from the original one via an appropriate parameter perturbation; and (iii) provide a formal certificate on the quality of the approximation as an error bound, computed as an over-approximation of the reachable set of the perturbed model. Finally, we apply approximate differential equivalences to study the effect of parametric tolerances in models of symmetric electric circuits. 
\end{abstract}

\input{intro}

\input{idol}

\input{technical} 
\input{bounds}
\input{caseStudies}

\section{Conclusion}
Reasoning about quantitative properties approximately can represent an effective way of taming the complexity of real systems. Here we have considered ordinary differential equations (ODEs) with polynomial derivatives. We developed notions of equivalence as a relaxation of their exact counterparts, allowing the derivatives of related ODE variables to vary up to a desired tolerance. Our algorithmic approach can be useful to systematically discover quasi-symmetries in  situations such as those presented in our case study. In future work, it would be also possible to integrate other bounding techniques, such as~\cite{tac15} which lacks an automatic synthesis of a reference model but can offer a tradeoff between tightness of the bound and computation cost in its derivation.  

\section*{Acknowledgement}
Luca Cardelli is partially funded by a Royal Society Research Professorship. Mirco Tribastone is supported by a DFG Mercator Fellowship (SPP 1593, DAPS2 Project). Max Tschaikowski is supported by a Lise Meitner Fellowship funded by the Austrian Science Fund (FWF) under grant number M 2393-N32 (COCO).

\bibliographystyle{splncs03}
\bibliography{qest}





\end{document}

%% file: macrosLNCS.tex
\pdfoutput=1

\usepackage{numprint}
\usepackage[english]{babel}

\newcommand{\h}{\mathcal{H}}

\newcommand{\RE}{\mathbb{R}}            

\usepackage{relsize}

\usepackage{pepa}
\def\equationautorefname~#1\null{Equation~(#1)\null}

\usepackage{listings}


%% file: intro.tex
\pdfoutput=1

\section{Introduction}

Ordinary differential equations (ODEs) are a prominent model of dynamical systems across many branches of science and engineering, and have enjoyed increasing popularity in computer science, for instance, in computational systems biology~\cite{Danos200469,Pedersen:2010aa,arand2012vivo}, as an approximation to large-scale Markov models and as the laws of continuous motion in hybrid systems~\cite{DBLP:conf/cav/BogomolovFGLPW12}. This has motivated techniques for the \emph{comparison} and \emph{minimization} of ODEs based on behavioral relations, along the lines of  other foundational quantitative models of computation, e.g.~\cite{Larsen19911}. Here we consider \emph{differential equivalence}~\cite{popl16}, recently developed as an equivalence over ODE variables yielding a quotient that preserves the dynamics of the original one. However, differential equivalences (reviewed in Section~\ref{sec_background}) are exact, hence highly sensitive to parameter values and initial conditions. This may hinder their practical usability in some applications domains, for instance due to parameter uncertainty arising from finite-precision measurements in biology or the tolerances of electric components in electrical engineering.

Our objective is to develop approximate variants of differential equivalence (in Section~\ref{sec:ade}). We study models with derivatives given by multivariate polynomials (over the ODE variables) of any degree, thus restricting the scope of~\cite{popl16}. However we remark that this is still quite a generous class since it includes chemical reaction networks (CRNs) with mass-action kinetics and linear/affine systems, thus covering, e.g., continuous-time Markov chains through their Kolmogorov equations.

Considering polynomial derivatives allows us to introduce a notion of equivalence that just concerns the ODE ``syntax'' (while the nonlinear class of ODEs of~\cite{popl16} required symbolic SMT-based checks). Our main idea is to consider a threshold parameter $\varepsilon \geq 0$, which intuitively captures perturbations in polynomials coefficients. This allows relating ODE variables that would be distinct otherwise. Like in other established approaches such as behavioral pseudometrics (e.g.,~\cite{DBLP:conf/cav/AbateBCK15,worrell01:approximate}), $\varepsilon = 0$ corresponds to an exact differential equivalence of~\cite{popl16}. In addition to defining criteria for approximate differential equivalences, we provide an algorithm for obtaining the largest one. This is done via partition refinement, computing the coarsest refinement of a given initial partition of ODE variables for a given ``structural'' tolerance $\varepsilon$.

A quotient ODE system can be constructed from a \emph{reference model}, obtained through a perturbation of the coefficients of the original model which makes the given approximate differential equivalence an exact one. By considering a metric (the Euclidean norm) to measure the degree of perturbation, the reference model is the one which minimizes such perturbation. This can be done efficiently by solving an optimization problem which runs polynomially with the size of the ODE system~\cite{KOZLOV1980223}. This approach is analogous to optimal approximate lumping for Markov chains (e.g.,~\cite{E:2008aa}), although our theory can be applied to other choices of reference models.


The bound of the error produced by the reference model with respect to the original system can be computed by studying the reachable set of the reference model from an uncertain set of initial conditions that covers the applied perturbation. Since the reference model subsumes any behavior of the quotient, the bound formally relates the quotient model to the original model. Section~\ref{sec:error} presents a bound which relies on a linearization of the reference model (which can be efficiently computed in the case of polynomial ODE systems). First, we bound the reachable set of the linearized model using closed form solutions, similarly to~\cite{DBLP:conf/emsoft/LalP15}. Then, we provide a conservative condition (i.e., an over-approximation) that ensures that the linearized model describes the original nonlinear behavior dynamics well. Our bound is given in terms of an $\varepsilon$-$\delta$ argument (similar in spirit to the ones routinely used in calculus). Informally, it states the following: for any choice of the structural tolerance $\varepsilon$, there exist a degree of perturbation $\delta$ and an \emph{amplifier} $\lambda$ such that, for any ODE system obtained by applying a perturbation to the reference model of at most $\delta$, at all time points the difference between the solution of the reference model and the perturbed one is at most $\lambda$ times the perturbation.

Being based on a linearization, it is perhaps not surprising that the as-computed $\delta$ will account only for \emph{small} perturbations of the parameters. Yet numerical experiments in Section~\ref{sec:numerical} show that these can be enough to explain quasi-symmetric behavior due to parametric tolerances in components of real electric network designs~\cite{DBLP:journals/tvlsi/RosenfeldF07}.
By comparing to over-approximation techniques supported by state-of-the-art tools C2E2~\cite{DBLP:conf/cav/FanQM0D16}, CORA~\cite{Althoff2015a,Althoff2013a} and Flow$^\ast$~\cite{DBLP:conf/cav/ChenAS13}, we show that our bounding technique can complement them in that it can scale to larger systems while being more conservative in the size of the initial uncertain set that it supports.

\paragraph*{Further related work.} Differential equivalence is promising when the ODE system is composed of several identical subsystems that depend on some common context~\cite{DBLP:journals/tcs/TschaikowskiT14}. It is  related but not comparable to bisimulation for differential systems~\cite{1582235,DBLP:journals/tcs/IslamMBCFGSG15} since it partitions ODE variables rather than the state space. Likewise, it complements~\cite{DBLP:conf/fossacs/Boreale17} that captures nonlinear relations between ODE variables but does not enjoy a polynomial time algorithm like~\cite{cttvPNAS}.

A classic approximation approach relies on Lyapunov-like functions~\cite{DBLP:conf/cav/MajumdarZ12,Duggirala:2013:VAM:2555754.2555780} known from stability theory of ODEs. However, for nonlinear systems the automatic computation of Lyapunov-like functions remains a challenging task. Restricting to special classes of Lyapunov-like functions (e.g., sum-of-squares polynomials~\cite{1582235}) leads to efficient construction algorithms which may provide tight bounds, but existence is not guaranteed. On the other hand, approximations with differential inequalities~\cite{tac15} can be computed efficiently, but may be loose. Abstraction, supported by CORA and Flow$^\ast$,  locally approximates the nonlinear model by a multivariate polynomial or an affine system, see \cite{Asarin2003,DBLP:conf/cav/ChenAS13} and references therein. Similarly in spirit we linearize across a reference trajectory. A closer approach to ours is discussed in~\cite{DBLP:conf/cav/FanQM0D16} and supported by the corresponding tool C2E2. It combines local Lyapunov-like functions and techniques based on sensitivity analysis~\cite{DBLP:conf/emsoft/LalP15}. Our bound is however different because the nonlinear part is bounded analytically by restricting to polynomial derivatives.

More in general, research on approximate quotients of ODE systems spans many disciplines. In chemistry, it can be traced back to Kuo and Wei~\cite{KuoWei:1969}. They studied monomolecular reaction networks, which give rise to affine ODE systems. The approximation consists in \emph{nearly exact lumping}, i.e., a linear transformation of the state space that would be exact up to a perturbation of the parameters (hence we are similar in spirit). The approximation, however, only applies when the transition matrix underlying the linear system is diagonalizable. Li and Rabitz extend approximate lumping to general CRNs~\cite{LI1990977}, but an explicit error bound is not given. In a similar vein, approximate quotients in ecology have been studied from the point of view of finding a reduced ODE system whose derivatives are as close as possible (in norm) to the derivatives of the original ODE system, where the 0-distance induce the exact quotient~\cite{IWASA01011989}. The justification that variables underlying similar ODEs have nearby solutions is grounded on Gronwall's inequality which is also at the basis of more recent quotient constructions~\cite{dsn13}, which however are not algorithmic, unlike in this paper.

\paragraph*{Notation and basic definitions.} We denote the infinity norm $\lVert \cdot \rVert_\infinity$ by $\lVert \cdot \rVert $, while $\lVert \cdot \rVert_2$ is the Euclidian norm.
Whenever convenient, for a given partition of variables $\mathcal{H}$, we write $H = \{\syn{x}_{H,1},\ldots,\syn{x}_{H,|H|}\}$ for any $H \in \mathcal{H}$. We denote by $\psi[t/s]$ the term that arises by replacing each occurrence of $t$ in $\psi$ by $s$.  Let $S$ be a set, and $\mathcal{H}_1$, $\mathcal{H}_2$ two partitions of $S$. Then, $\mathcal{H}_1$ is a \emph{refinement} of $\mathcal{H}_2$ if for any block $H_1 \in \mathcal{H}_1$ there exists a block $H_2 \in \mathcal{H}_2$ such that $H_1 \subseteq H_2$. For any partition $\h$ of $S$, let $\sim_\h$ denote the unique equivalence relation with $\h = S / {\sim_\h}$. The transitive closure of a relation $\sim$ is denoted by $\sim^\ast$.

%% file: idol.tex

\pdfoutput=1

\section{Background}\label{sec_background}

Throughout the paper we consider a polynomial initial value problem (PIVP) over the set of ODE variables $\dvars = \{ x_1, \ldots, x_n \}$. It is defined by the ODEs $\dot{x}_i = q_i$,  $1 \leq i \leq n$, where $q_i$ is a multivariate polynomial over $\dvars$. The initial condition of the PIVP is given by $\sigma : \dvars \rightarrow \mathbb{R}$; $x_i(t)$ denotes the unique solution for variable $x_i$ at time point $t$ starting from $x_i(0) = \sigma(x_i)$. We consider PIVPs that do not exhibit finite explosion times, i.e., whose solutions do not have a singularity at any finite point in
time. This property is shared by the vast majority of practical models and can be efficiently checked via numerical ODE solvers.

A polynomial $q_i$ is given in the normal form if each monomial $x^\alpha \equiv \prod_{x_i \in \dvars} x_i^{\alpha_{x_i}}$, where $\alpha \in \mathbb{N}_0^\dvars$ is a multi-index, appears in $q_i$ at most once. The normal form of a polynomial $q_i$ is denoted by $\nf(q_i)$. Without loss of generality, we assume that the polynomials $q_i$ of a PIVP $\dot{x} = q(x)$, where $q = (q_i)_{x_i \in \dvars}$, are given in normal form.  For a polynomial $q_i$ in normal form with variables in $\dvars$, let $c(q_i,x^\alpha)$ denote the coefficient of the monomial $x^\alpha$, where $\alpha \in \mathbb{N}_0^\dvars$. 

\begin{example}\label{ex_re}
We use the following ODE system, with variables $\dvars = \{ x_1, x_2, x_3\}$, as a running example.
\begin{align}\label{ex:ctmc}
\dot{x}_1  & = - 4.00 x_1 + x_2 + x_3 &
\dot{x}_2 & = 1.99 x_1 - x_2  &
\dot{x}_3 & = 2.01 x_1 - x_3
\end{align}
\end{example}

In~\cite{popl16} two variants of differential equivalence were introduced for \idol{}, a class of nonlinear ODE systems covering derivatives more general than polynomials. Here we find it convenient to restate them for a PIVP. (The proofs for this correspondence are straightforward hence we omit them.)

We begin with \emph{backward differential equivalence} (BDE), which relates variables that have the same solutions at all time points.
The definition of BDE for PIVP makes pairwise comparisons between the coefficients of any two variables  in the same equivalence class.

\begin{definition}[BDE]\label{def_bde}
Fix a PIVP, a partition $\h$ of $\dvars$ and write $x_i \ebnt x_j$ if all coefficients of the following polynomial are zero,
$$\wpb := (q_i - q_j) \big[ x_{H',1} \big / x_{H'}, \ldots, x_{H',|H'|} \big / x_{H'} \!:\! H' \in \!\mathcal{H} \big]$$i.e., when
\begin{equation}\label{eq:bdecon}
\sum_{\alpha \in \mathbb{N}_0^{\dvars}} | c(\wpb,x^\alpha) | = 0 .
\end{equation}
A partition $\h$ is a BDE if $\h = \dvars / {(\ebt \cap \sim_\h)}$.
\end{definition}

Essentially, establishing that a candidate partition is a BDE consists in comparing the coefficients of the monomials of the ODEs of related variables, up to the natural equivalence class induced on monomials by the equivalence relation through $\wpb$ --- for instance, the partition $\{ \{ x_1\}, \{ x_2, x_3 \} \}$ will equate the monomials $x_1x_2$ and $x_1x_3$.
Then, for any two variables in the same block it must hold that the differences between the coefficients of the same monomials (modulo the induced equivalence class) are zero.


\begin{example}\label{ex_re_bde}
In our running example let us consider the partition of variables $\h = \{ H_1, H_2 \}$, with $H_1 = \{x_1\}$ and $H_2 = \{x_2, x_3\}$. Then $\h$ is not a BDE because
$$\wp_{2,3}^\h = -0.02 x_1 \qquad \text{and} \qquad c(\wp_{2,3}^\h,x_1) = -0.02 \neq 0 . $$
\end{example}

\emph{Forward differential equivalence} (FDE) identifies a partition that induces a quotient ODE that tracks sums of variables in each equivalence class. For instance, for any initial condition we have that  $\{ \{ x_1 \}, \{ x_2, x_3 \}\}$ is an FDE for (\ref{ex:ctmc}) because we can find an ODE system for $x_2 + x_3$:
\begin{align*}
\dot{x}_1  & = - 4.00 x_1 + (x_2 + x_3) &
\dot{(x_2 + x_3)}  & = 4.00 x_1 - (x_2 + x_3)
\end{align*}
The change of variable $x_{23} = x_2 + x_3$ gives us the quotient ODE $\dot{x}_1 = - 4.00 x_1 + x_{23}$, $\dot{x}_{23} = 4.00 x_1 - x_{23}$. From this we conclude that the solution satisfies $x_{23}(t) = x_2(t) + x_3(t)$ for all time $t$ if this holds for the initial condition, i.e., $x_{23}(0) = x_2(0) + x_3(0)$.

For a PIVP, FDE can be checked by requiring that the evaluation of the polynomial that represents the quotient derivative for an equivalence class is invariant with respect to a redistribution of the values of any two variables within that equivalence class.

\begin{definition}[FDE]\label{def_fde}
Fix a PIVP, a partition $\h$ of $\dvars$ and write $x_i \efnt x_j$ if all coefficients of the polynomial $\sum_{H \in \h} \wpf$ are zero, where
$$\wpf := \sum_{x_k \in H} q_k - \sum_{x_k \in H} q_k[x_i / s(x_i + x_j), x_j / (1 - s) (x_i + x_j)]$$
That is, when
\begin{equation}\label{eq:fdecon}
\sum_{k = 1}^m \sum_{\alpha \in \mathbb{N}_0^{\dvars \dot \cup \{s\}}} | c(\wp_{i,j}^{H_k},x^\alpha) | = 0. \
\end{equation}
$\h = \{H_1,\ldots,H_m\}$ is an FDE when $\h = \dvars /{(\eft \cap \sim_\h)}$.
\end{definition}

%% file: technical.tex

\pdfoutput=1

\newcommand{\frt}{\mathfrak{t}}

\section{Approximate Differential Equivalences}\label{sec:ade}



\paragraph*{Definitions.} Approximate differential equivalence relaxes the equality conditions (\ref{eq:bdecon})-(\ref{eq:fdecon}) of Definition~\ref{def_bde} and~\ref{def_fde} to inequalities with respect to a tolerance level $\varepsilon$.

\begin{definition}[Approximate BDE]
Fix a PIVP, a partition $\h = \{H_1,\ldots,H_m\}$ of $\dvars$, and $\varepsilon \geq 0$. We write $x_i \bnt x_j$ if $\sum_{\alpha \in \mathbb{N}_0^{\dvars}} | c(\wpb,x^\alpha) | \leq \varepsilon$, where $\wpb$ is as in Definition~\ref{def_bde}. A partition $\h$ is an $\varepsilon$-BDE if $\h = \dvars / {(\bt \cap \sim_\h)}$.
\end{definition}

\begin{definition}[Approximate FDE]
Fix a PIVP, a partition $\h = \{H_1,\ldots,H_m\}$ of $\dvars$, and $\varepsilon \geq 0$. We write $x_i \fnt x_j$ if $\sum_{k = 1}^m \sum_{\alpha \in \mathbb{N}_0^{\dvars \dot \cup \{s\}}} | c(\wp_{i,j}^{H_k},x^\alpha) | \leq \varepsilon$, where $\wpf$ is as in Definition~\ref{def_fde}. A partition $\h$ is an $\varepsilon$-FDE when $\h = \dvars /{(\ft \cap \sim_\h)}$.
\end{definition}

Setting $\varepsilon = 0$ recovers the exact counterparts in both cases. That is, $\h$ is an BDE (resp., FDE) partition if and only if $\h$ is a 0-BDE (resp., 0-FDE) partition.The two approximate differential equivalences are not comparable since their exact counterparts are not~\cite{popl16}. Since these two notions have similar structure
in the rest of this paper we will illustrate only approximate BDE using small examples. Instead, both notions will be discussed in more detail for the numerical evaluation of Section~\ref{sec:numerical}.


\begin{example}\label{ex_mc0}
Let us consider our running example~(\ref{ex:ctmc}). Then, the partition $\big\{ \{x_1\}, \{x_2, x_3\} \big\}$ is a $0.02$-BDE partition, as can be easily seen from Example~\ref{ex_re_bde}.
\end{example}

The next two results show the existence of largest approximate differential equivalences and of a partition-refinement algorithm to compute it.

\begin{theorem}\label{thm_ex_coarsest}
Fix a PIVP, a partition $\mathcal{G}$ of $\dvars$, and $\varepsilon \geq 0$. Then, there exists a unique coarsest $\varepsilon$-FDE ($\varepsilon$-BDE) partition refining $\mathcal{G}$.
\end{theorem}


\begin{algorithm}[t!]
\caption{Template partition refinement algorithm for the computation of the coarsest $\varepsilon$-FDE/$\varepsilon$-BDE partition that refines a given initial partition $\mathcal{G}$.}\label{algorithm_part}
\begin{algorithmic}
\REQUIRE A PIVP over variables $\dvars$, a partition $\mathcal{G}$ of $\dvars$, a threshold $\varepsilon \geq 0$, and a mode $\chi \in \{F,B\}$.

\STATE $\h$ $\longleftarrow$ $\mathcal{G}$
\WHILE{\TRUE}
    \STATE $\h' \longleftarrow$ $S / (\sim^{\chi^\ast}_{\h,\varepsilon} \cap \sim_\mathcal{H})$

    \IF{$\h' = \h$}
        \RETURN $\h$
    \ELSE
        \STATE $\h \longleftarrow \h'$
    \ENDIF
\ENDWHILE
\end{algorithmic}
\end{algorithm}

\begin{theorem}\label{thm_cofl}
Fix a PIVP,  a partition $\mathcal{G}$ of $\dvars$, and $\varepsilon \geq 0$. Then, Algorithm~\ref{algorithm_part} computes the coarsest $\varepsilon$-FDE ($\varepsilon$-BDE) that refines $\mathcal{G}$ if $\chi = F$ ($\chi = B$).
\end{theorem}

We now study how efficiently the conditions for approximate differential equivalence can be computed. The notions are defined with respect to the coefficients of the polynomials $\wpf$ and $\wpb$ and thus require the computation of their normalization. In the case of $\varepsilon$-FDE, this yields exponential complexity due to term replacement. To see this, consider for instance the PIVP $\dot{x}_1 = x_2^{k}, \dot{x}_2 = x_1^{k}$, for some $k >0$. Then, for $\h = \big\{ \{x_1, x_2\} \}$, the term $q_1[x_1 / s(x_1 + x_2), x_2 / (1 - s) (x_1 + x_2)]$
will be of size $\mathcal{O}(2^k)$. This stands in stark contrast to $\varepsilon$-BDE, where the conditions involve a difference between polynomials terms with no term rewritings. This discussion can be formalized as follows.

\begin{theorem}\label{thm_complex}
There exists a polynomial $\Pi$ such that, under the assumptions of Theorem~\ref{thm_cofl}, the number of steps done by Algorithm~\ref{algorithm_part} is $\mathcal{O}\big(\Pi(2^{d} \cdot p)\big)$ if $\chi = F$ and $\mathcal{O}\big(\Pi(p)\big)$ if $\chi = B$, respectively, where $d$ is the maximum degree of the polynomial and $p$ is the number of monomials present in the PIVP.
\end{theorem}
In practice, $d$ is usually not large. For example, mass-action CRNs feature ODEs with degree-two polynomials because in nature at most two species interact in every reaction. An experimental comparison between the reduction runtimes of $\varepsilon$-FDE and $\varepsilon$-BDE will be presented in Section~\ref{sec:numerical}. We also remark that since $0$-FDE/BDE coincides with FDE/BDE, the above result provides a complexity bound for a subclass of ODE systems considered in~\cite{popl16}.

Other computational considerations motivate the choice of the definitions of approximate differential equivalence given in this paper. Another natural definition could have involved the computation of  the maximal distance between derivatives ``semantically'', i.e., under all possible evaluations within a given domain of interest. For example, consider the PIVP $\dot{x}_1 = x_1^3 - x_2$, $\dot{x}_2 = x_1 - x_2^3$. Establishing that $\{ \{ x_1, x_2 \} \}$ is an $\varepsilon$-BDE would require checking that the difference between the derivatives satisfies
\begin{equation}\label{eq:optimization}
\lvert \dot{x}_1 - \dot{x}_2 \rvert = \lvert x_1^3 - x_1 + x_2^3 - x_2 \rvert \leq \varepsilon,
\text{for all~} 0 \leq x_1, x_2 \leq C
\end{equation}
for some finite $C > 0$ that represents some bounded domain where the trajectories are assumed to live. Since this example shows that this question is in general equivalent to solving a nonconvex optimization problem, we  infer that the problem is NP-hard~\cite{Pardalos1991}.

However it can be easily shown that our approximate differential equivalence, defined through the coefficients of the polynomials, corresponds to checks such as (\ref{eq:optimization}) in the following sense: If a partition $\h$ satisfies constraints similar to $(\ref{eq:optimization})$ with respect to some $\varepsilon > 0$, then there exists an $\varepsilon' > 0$ such that $\h$ is an $\varepsilon'$-FDE/BDE, and vice versa. The basic idea is to observe that a polynomial is the zero function if and only if its coefficients are all zero.

Finally, we remark that our structural/syntactic criteria can be used for PIVPs only. It is the lack of  analogous  conditions in the case of more general functions like minima or roots which prevents our approximate differential equivalences to be extended in a straightforward way to the full class of nonlinear ODEs of~\cite{popl16}.

\paragraph{Reference PIVP.} Given a partition of variables that represents an approximate differential equivalence, we construct a \emph{reference PIVP}  by finding a ``perturbation'' of the original PIVP --- i.e., a modification of the initial condition $\sigma$ and the coefficients present in $q_1,\ldots,q_n$ --- which ensures that that very partition becomes an exact differential equivalence. On this reference PIVP one can use the quotienting algorithms for FDE/BDE developed in~\cite{popl16} (and not restated here formally for brevity). Therefore, the as-obtained quotient represents an approximate reduction of the original PIVP.

We obtain the desired perturbation by treating the original initial conditions and polynomial coefficients uniformly as initial conditions on an
\emph{extended} PIVP where every coefficient is parameterized and turned into a new ODE variable.

\begin{definition}
The parameterization of a polynomial $q_i$ in normal form with variables $\dvars$ is denoted by $\hat{q}_i$ and arises from $q_i$ by replacing, for each $\alpha \in \mathbb{N}_0^\dvars$, the constant $c(q_i,x^\alpha)$ with the parameter $\frc(\hat{q}_i,x^\alpha)$.
\end{definition}

\begin{example}\label{ex_hatqnf}
The polynomials $q_2  = 1.99 x_1 - x_2$ and $q_3 = 2.01 x_1 - x_3$ from Example~\ref{ex_re} give rise to the parameterized polynomials $\hat{q}_2 = \frc(\hat{q}_2,x_1) x_1 + \frc(\hat{q}_2,x_2) x_2$ and $\hat{q}_3 = \frc(\hat{q}_3,x_1) x_1 + \frc(\hat{q}_2,x_3) x_3$, respectively.
\end{example}

\begin{definition}[Extended PIVP]\label{def_conf}
For a PIVP $\calp$ with variables $\dvars$, set $\Theta = \{ \frc(\hat{q}_i,x^\alpha) \mid 1 \leq i \leq n, \alpha \in \mathbb{N}_0^\dvars \}$. Its extended version $\hat{\calp}$ has variables $\dvars \cup \Theta$ and is given by $\dot{x}_i = \hat{q}_i$ and $\dot{\frc}(\hat{q}_i,x^\alpha) = 0$, where $x_i \in \dvars$ and $\alpha \in \mathbb{N}_0^\dvars$. For a given $\hs \in \RE^{\dvars \cup \Theta}$, let $\hat{\calp}(\hs)$ denote the PIVP which arises from $\hat{\calp}$ by replacing each $v \in \dvars \cup \Theta$ by the corresponding real value $\sigma(v) \in \RE$ in $\hat{\calp}$. In particular, let $\hs_0 \in \RE^{\dvars \cup \Theta}$ be such that $\calp(\sigma) = \hat{\calp}(\hs_0)$.
\end{definition}

\begin{example}\label{ex_hatcalp}
If $\calp$ is the PIVP from Example~\ref{ex_re}, its extended version $\hat{\calp}$ is
\begin{align*}
\dot{x}_1 & = \frc(\hat{q}_1,x_1) x_1 + \frc(\hat{q}_1,x_2) x_2 + \frc(\hat{q}_1,x_3) x_3, & \dot{\frc}(\hat{q}_1,x_i) & = 0, \quad i =1,2,3, \\
\dot{x}_2 & = \frc(\hat{q}_2,x_1) x_1 + \frc(\hat{q}_2,x_2) x_2,  & \dot{\frc}(\hat{q}_2,x_i) & = 0, \quad i =1,2,3, \\
\dot{x}_3 & = \frc(\hat{q}_3,x_1) x_1 + \frc(\hat{q}_3,x_2) x_2,  & \dot{\frc}(\hat{q}_3,x_i) & = 0, \quad i =1,2,3.
\end{align*}
The corresponding $\hs_0$ satisfies $\hs_0(x_i) = \sigma(x_i)$ for $1 \leq i \leq 3$ and
\begin{align*}
\hs_0\big(\frc(\hat{q}_1,x_1)\big) & = -4.00, & \hs_0\big(\frc(\hat{q}_1,x_2)\big) & = 1.00, & \hs_0\big(\frc(\hat{q}_1,x_3)\big) & = 1.00, \\
\hs_0\big(\frc(\hat{q}_2,x_1)\big) & = 1.99, & \hs_0\big(\frc(\hat{q}_2,x_2)\big) & = -1.00, & \\
\hs_0\big(\frc(\hat{q}_3,x_1)\big) & = 2.01, & \hs_0\big(\frc(\hat{q}_3,x_2)\big) & = -1.00. &
\end{align*}
\end{example}

The following is needed for the definition of the reference PIVP.

\begin{definition}
Given constant free polynomial $\hat{\wp}$ (i.e., such that $\hat{\wp}(0)=0$) and $\Xi \subseteq \dvars \cup \Theta \cup \{s\}$, let $\frt(\hat{\wp},x^\alpha,\Xi)$ denote the coefficient term of $x^\alpha$ in $\nf(\hat{\wp},\Xi)$, where $\alpha \in \mathbb{N}_0^{\Xi}$ and $\nf(\hat{\wp},\Xi)$ is the normal form of $\hat{\wp}$ where variables outside $\Xi$ are treated as parameters.
\end{definition}

\begin{example}
With $\hat{q}_2$ and $\hat{q}_3$ as in Example~\ref{ex_hatqnf} and $\Xi = \{x_1,x_2,x_3\}$, the normal form $\nf(\hat{q}_2 - \hat{q}_3,\Xi)$ is given by $(\frc(\hat{q}_2,x_1) - \frc(\hat{q}_3,x_1)) x_1 + (\frc(\hat{q}_2,x_2) - \frc(\hat{q}_3,x_2)) x_2 $, while $\frt(\hat{q}_2 - \hat{q}_3,x_1,\Xi) = \frc(\hat{q}_2,x_1) - \frc(\hat{q}_3,x_1)$.
\end{example}


\begin{definition}\label{def_lin_const}
Given a PIVP with variables $\dvars$ and an $\varepsilon$-FDE partition $\h$ of $\dvars$, the set of linear constraints of $\h$ is given by
\begin{align}\label{eq:sys.fde}
\big\{ \frt(\tilde{\wp}^H_{i,j}, x^\alpha, \dvars \cup \{s\}) = 0 \mid \alpha \in \mathbb{N}_0^{\dvars \cup \{s\}}, H \in \h \text{ and } x_i \ {\sim_\h} \ x_j \big\}
\end{align}
with $\tilde{\wp}^H_{i,j} = \sum_{x_k \in H} \hat{q}_k - \sum_{x_k \in H} \hat{q}_k[x_i / s(x_i + x_j), x_j / (1 - s) (x_i + x_j)]$.

If $\h$ is an $\varepsilon$-BDE partition of $\dvars$, the corresponding set of linear constraints is
\begin{multline}\label{eq:sys.bde}
\big\{ \frt(\tilde{\wp}^\h_{i,j},x^\alpha, \dvars) = 0 \mid \alpha \in \mathbb{N}_0^\dvars, x_i \ {\sim_\h} \ x_j \big\}  \\
\mathop{\cup} \big\{ x_{i_j} - x_{i_{j+1}} = 0 \mid 1 \leq j \leq k-1 \text{ and } \{x_{i_1},\ldots,x_{i_k}\} \in \dvars / {\sim_\h} \big \} ,
\end{multline}
where $\tilde{\wp}^\h_{i,j} = (\hat{q}_i - \hat{q}_j) \big[ x_{H',1} \big / x_{H'}, \ldots, x_{H',|H'|} \big / x_{H'} \!:\! H' \in \!\mathcal{H} \big]$.
\end{definition}

\begin{example}\label{ex_lin_cons}
From Example~\ref{ex_re_bde}, we know that $\h = \{ \{x_1\}, \{x_2, x_3\}\}$ is a $0.02$-BDE partition of the PIVP~(\ref{ex:ctmc}). The set of linear constraints underlying $\h$ is given by $\frc(\hat{q}_2,x_1) - \frc(\hat{q}_3,x_1) = 0$ and $x_2 - x_3 = 0$.
\end{example}

\begin{remark} In line with its exact counterpart, an $\varepsilon$-BDE is ``useful'' under the further constraint that related variables have the same initial conditions in the reference model, as a necessary condition for having equal solutions at all time points. This translates into adding the constraints in (\ref{eq:sys.bde}) that perturbed initial conditions of related variables are equal. This leads, for instance, to the constraint $x_2 - x_3 = 0$ in the running example. For $\varepsilon$-FDE, instead, only constraints on the parameters $\Theta$ are made.
\end{remark}

\begin{theorem}\label{prop_lin}
Given a PIVP $\calp$ with variables $\dvars$, an $\varepsilon$-FDE/BDE partition $\h$ and a configuration $\hs \in \RE^{\dvars \cup \Theta}$ that satisfies~(\ref{eq:sys.fde})/(\ref{eq:sys.bde}), it holds that $\h$ is an FDE/BDE of $\hat{\calp}(\hs)$.
\end{theorem}

The linear system of constraints from Theorem~\ref{prop_lin} can be shown to be underdetermined, hence there are infinitely many perturbations that lead to an exact differential equivalence. This observation is an instance of the well-known fact that, in general, an approximate quotient is not unique. Here, we fix one candidate perturbation by assuming that nearby initial conditions yield nearby trajectories. This fact is asymptotically true due to Gronwall's inequality, as mentioned in Section~1.

We are interested in finding a configuration $\hs$ which satisfies the constraints of Theorem~\ref{prop_lin} and minimizes the distance $\lVert \hs - \hs_0 \rVert_2$. Mathematically, this corresponds to the optimization problem
\begin{align}\label{eq_optim}
\hso = \argmin_{\hs : \text{Eq. (\ref{eq:sys.fde})/(\ref{eq:sys.bde}) holds}} \lVert \hs - \hs_0 \rVert_2
\end{align}
Since the solution space of a linear system is convex, the Euclidian norm yields a convex quadratic program that can be solved in polynomial time~\cite{KOZLOV1980223}.

\begin{example}
Let us continue Example~\ref{ex_lin_cons} and assume that $\sigma(x_2) = \sigma(x_3)$. In such a case, it can be easily seen that $\hso$ and $\hs_0$ satisfy $\hso(\frc(\hat{q}_2,x_1)) = \hso(\frc(\hat{q}_3,x_1)) = \big(\hs_0(\frc(\hat{q}_2,x_1)) + \hs_0(\frc(\hat{q}_3,x_1))\big) / 2 = 2.00$ and coincide on all other parameters. In other words, the closest PIVP that enjoys an exact BDE relating $x_2$ and $x_3$ is given, as expected, by perturbing the coefficients $1.99$ and $2.01$ of (\ref{ex:ctmc}) to their average value, yielding:
\begin{align*}
\dot{x}_1  & = - 4.00 x_1 + x_2 + x_3 &
\dot{x}_2 & = 2.00 x_1 - x_2  &
\dot{x}_3 & = 2.00 x_1 - x_3
\end{align*}
\end{example}

The above discussions are summarized in the following.

\begin{theorem}\label{thm_main_1}
Given a PIVP, $\varepsilon \geq 0$, and an $\varepsilon$-FDE/BDE partition $\h$, the solution of~(\ref{eq_optim}) exists and can be computed in polynomial time.
\end{theorem}


\begin{figure}[tp!]
\begin{center}
    \input{figopt2.pdftex_t}
\end{center}
\caption{Given a PIVP $\calp$, a partition $\g$ of $\dvars$, and an $\varepsilon > 0$, the coarsest $\varepsilon$-FDE/BDE partition $\h$ that refines $\g$ is constructed. Afterwards, the solution $\hso$ of the optimization problem~(\ref{eq_optim}) is computed in Fig. 1(a). This allows to compute the $\varepsilon$-FDE/BDE quotient $\hat{\calp}(\hso)$ of $\h$. With this, $\lambda$ and $\delta$ from Theorem~\ref{thm_bound} are calculated. In the case the distance between $\hs_0$ and $\hso$ does not exceed $\delta$, the tight bounds of Theorem~\ref{thm_bound} can be applied and relate the trajectories of $\hat{\calp}(\hso)$ and $\hat{\calp}(\hs_0) = \calp(\sigma)$, as depicted in Fig. 1(b).}\label{fig_opt_and_ball}
\end{figure}

The solution of the optimization problem~(\ref{eq_optim}) stated in Theorem~\ref{thm_main_1} is informally depicted in Fig.~\ref{fig_opt_and_ball}a.

The reference PIVP is the extended, exactly reducible PIVP with the optimum initial condition $\hso$, i.e., $\hat{\calp}(\hso)$. Its ODE solution is called the \emph{reference trajectory}.

%% file: figopt2.pdftex_t
\pdfoutput=1

\begin{picture}(0,0)%
\includegraphics{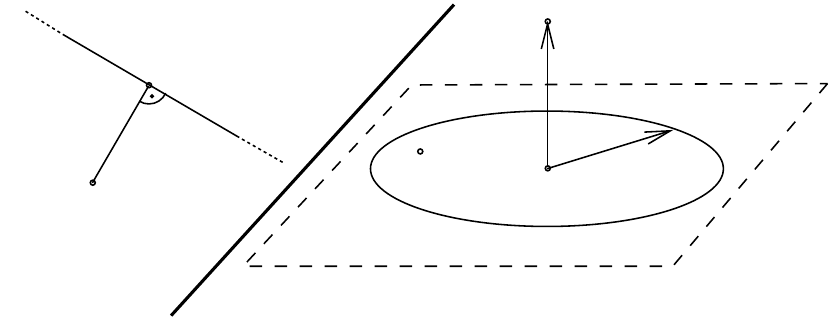}%
\end{picture}%
%
%
\setlength{\unitlength}{3947sp}%
\begingroup\makeatletter\ifx\SetFigFont\undefined%
\gdef\SetFigFont#1#2#3#4#5{%
  \reset@font\fontsize{#1}{#2pt}%
  \fontfamily{#3}\fontseries{#4}\fontshape{#5}%
  \selectfont}%
\fi\endgroup%
\begin{picture}(3984,1570)(103,-776)
\put(2920, 83){\makebox(0,0)[lb]{\smash{{\SetFigFont{8}{9.6}{\rmdefault}{\mddefault}{\updefault}{\color[rgb]{0,0,0}$\delta$}%
}}}}
\put(464,-219){\makebox(0,0)[lb]{\smash{{\SetFigFont{8}{9.6}{\rmdefault}{\mddefault}{\updefault}{\color[rgb]{0,0,0}$\hat{\sigma}_0$}%
}}}}
\put(841,413){\makebox(0,0)[lb]{\smash{{\SetFigFont{8}{9.6}{\rmdefault}{\mddefault}{\updefault}{\color[rgb]{0,0,0}$\hat{\sigma}_\ast$}%
}}}}
\put(486,614){\makebox(0,0)[lb]{\smash{{\SetFigFont{8}{9.6}{\rmdefault}{\mddefault}{\updefault}{\color[rgb]{0,0,0}Solution set of Eq. (5) / (6)}%
}}}}
\put(527,-33){\rotatebox{59.0}{\makebox(0,0)[lb]{\smash{{\SetFigFont{8}{9.6}{\rmdefault}{\mddefault}{\updefault}{\color[rgb]{0,0,0}distance}%
}}}}}
\put(2659,-154){\makebox(0,0)[lb]{\smash{{\SetFigFont{8}{9.6}{\rmdefault}{\mddefault}{\updefault}{\color[rgb]{0,0,0}$\hat{\sigma}_\ast$}%
}}}}
\put(2144,-41){\makebox(0,0)[lb]{\smash{{\SetFigFont{8}{9.6}{\rmdefault}{\mddefault}{\updefault}{\color[rgb]{0,0,0}$\hat{\sigma}_0$}%
}}}}
\put(2822,618){\makebox(0,0)[lb]{\smash{{\SetFigFont{8}{9.6}{\rmdefault}{\mddefault}{\updefault}{\color[rgb]{0,0,0}$\varepsilon$-FDE/BDE quotient}%
}}}}
\put(3631,-274){\makebox(0,0)[lb]{\smash{{\SetFigFont{8}{9.6}{\rmdefault}{\mddefault}{\updefault}{\color[rgb]{0,0,0}$\mathbb{R}^{\mathcal{S} \cup \Theta}$}%
}}}}
\put(2297,-713){\makebox(0,0)[lb]{\smash{{\SetFigFont{9}{10.8}{\rmdefault}{\mddefault}{\updefault}{\color[rgb]{0,0,0}Figure 1(b)}%
}}}}
\put(118,-713){\makebox(0,0)[lb]{\smash{{\SetFigFont{9}{10.8}{\rmdefault}{\mddefault}{\updefault}{\color[rgb]{0,0,0}Figure 1(a)}%
}}}}
\end{picture}%

%% file: bounds.tex

\pdfoutput=1

\section{Error Bounds}\label{sec:error}

The objective of this section is to provide a tight bound on the difference between the solution of the original PIVP and the reference. More specifically, we will show how to compute two values $\delta > 0$  and $\lambda > 0$ such that for all initial conditions $\hs_1 \in \RE^{S \cup \Theta}$ with $\lVert x^{\hs_1}(0) - x^{\hso}(0) \rVert \leq \delta$, it holds that $\max_{0 \leq t \leq \hat{\tau}} \lVert x^{\hs_1}(t) - x^{\hso}(t) \rVert \leq \lambda \lVert x^{\hs_1}(0) - x^{\hso}(0) \rVert$, where $x^{\hs}$ denotes the solution underlying $\hat{\calp}(\hs)$ and $\hat{\tau} > 0$ is a previously fixed finite time horizon.

The quantity $\delta$ gives the size of the ball around the initial condition $\hso$ of the reference PIVP, whereas $\lambda$ is the \emph{amplifier} that relates the maximum distance between trajectories to the distance between the initial conditions. Therefore, if the initial condition of the original PIVP $\hat{\calp}(\hs_0)$ falls within the prescribed $\delta$ ball, then the above statement will provide a formal bound of the error made in approximating the original PIVP $\hat{\calp}(\hs_0)$ with the reference PIVP. This idea is visualized in Fig.~\ref{fig_opt_and_ball}(b).

Let us recall the notion of Jacobian matrix.

\begin{definition}
Given an extended PIVP with variables $\dvars \cup \Theta$, the entries of the Jacobian matrix $A = (A_{i,j})_{x_i, x_j \in \dvars \cup \Theta}$ are given by $A_{i,j} = \partial_{\syn{x}_j} \hat{q}_i$, where $\partial_{\syn{x}}$ denotes the partial derivative with respect to $\syn{x}$.
\end{definition}
Let $A(t) \in \RE^{\dvars \cup \Theta \times \dvars \cup \Theta}$ denote the Jacobian obtained by plugging in the reference trajectory $x^{\hso}(t)$. We will need the following result from the theory of ODEs.

\begin{theorem}\label{thm_linear}
There exists a family of matrices $\Lambda(t_0, t_1)$, with $0 \leq t_0 \leq t_1 \leq \hat{\tau}$, such that the solution of $\dot{y}(t) = A(t) y(t)$, where $y(t_0) = y_0$ and $t_0 \leq t \leq \hat{\tau}$, is given by $y(t) = \Lambda(t_0, t) y_0$ for all $t_0 \leq t \leq \hat{\tau}$.
\end{theorem}

This is needed in the following.

\begin{theorem}\label{thm_bound}
Consider an extended PIVP $\hat{\calp}$ with variables $\dvars \cup \Theta$ and define $\lambda_0 = \max_{0 \leq t \leq \hat{\tau}} \lVert \Lambda(0,t)\rVert$ and $\lambda_1 = \max_{0 \leq t_0 \leq t_1 \leq \hat{\tau}} \lVert \Lambda(t_0,t_1)\rVert$. Further, define the remainder function $r : [0;\hat{\tau}] \times \mathbb{R}^{\dvars \cup \Theta} \to \mathbb{R}^{\dvars \cup \Theta}$ via
\begin{align*}
r(t, x - x^{\hso}(t)) = \hat{q}(x) - \hat{q}(x^{\hso}(t)) - A(t) (x - x^{\hso}(t))
\end{align*}
and let $0 \leq d_2,d_3,\ldots $ be such that $\lVert r(t,y) \rVert \leq \sum_{k = 2}^{\deg(\hat{\calp})} d_k \lVert y \rVert^k$ for all $y \in \mathbb{R}^{\dvars \cup \Theta}$ and $0 \leq t \leq \hat{\tau}$. Then, with $\lambda = 2 \lambda_0$, for any $x^{\hs_1}(0) \in \mathbb{R}^{\dvars \cup \Theta}$, it holds that
\[
\lVert x^{\hs_1}(0) - x^{\hso}(0) \rVert \leq \delta \ \Rightarrow \ \max_{0 \leq t \leq \hat{\tau}} \lVert x^{\hs_1}(t) - x^{\hso}(t) \rVert \leq \lambda \lVert x^{\hs_1}(0) - x^{\hso}(0) \rVert
\]
whenever $\delta > 0$ satisfies $\sum_{k = 2}^{\deg(\hat{\calp})} d_k (2 \lambda_0 \delta)^{k-1} \leq (2 \lambda_1 \hat{\tau})^{-1}$.
\end{theorem}
Theorem~\ref{thm_bound} provides a bound on the difference $x^{\hs_1}(t) - x^{\hso}(t)$ in terms of the initial perturbation $x^{\hs_1}(0) - x^{\hso}(0)$ if the latter is sufficiently small, i.e., does not exceed $\delta$. We wish to point out that the maximal $\delta$ satisfying $\sum_{k = 2}^{\deg(\hat{\calp})} d_k (2 \lambda_0 \delta)^{k-1} \leq (2 \lambda_1 \hat{\tau})^{-1}$ is a root of a polynomial in one variable and thus can be efficiently approximated from below via Newton's method. Instead, the assumption $\lVert r(s,y) \rVert \leq \sum_{k = 2}^{\deg(\hat{\calp})} d_k \lVert y \rVert^k$ on the remainder function $r$ states essentially that, for any $k \geq 2$, the sum of all $k$-th order derivatives of $r$ are bounded by $d_k$ along the reference trajectory $x^{\hso}$.

The next result shows that the bound of Theorem~\ref{thm_bound} is tight and relies on the sharp bound available in the special case of linear systems (i.e., $\deg(\hat{\calp}) = 1$) as discussed in~\cite{Donze2007}. 
\begin{theorem}\label{thm_tight}
If an extended PIVP $\hat{\calp}$ satisfies $\deg(\hat{\calp}) = 1$ and $\lambda = 2 \lambda_0$, it holds that
\[
\max_{0 \leq t \leq \hat{\tau}} \lVert x^{\hs_1}(t) - x^{\hso}(t) \rVert \leq \frac{\lambda}{2} \lVert x^{\hs_1}(0) - x^{\hso}(0) \rVert
\]
for any $x^{\hs_1}(0) \in \mathbb{R}^{\dvars \cup \Theta}$. The bound is tight in the sense that there exist $0 \leq t \leq \hat{\tau}$ and $x^{\hs_1}(0) \in \mathbb{R}^{\dvars \cup \Theta}$ such that $\lVert x^{\hs_1}(t) - x^{\hso}(t) \rVert = \frac{\lambda}{2} \lVert x^{\hs_1}(0) - x^{\hso}(0) \rVert$.
\end{theorem}
Note that the amplifier in Theorem~\ref{thm_bound} is twice as large as the amplifier in Theorem~\ref{thm_tight}. This is because the proof of Theorem~\ref{thm_bound} has to estimate nonlinear terms present in the remainder function $r$. More importantly, Theorem~\ref{thm_tight} shows that the amplifier of Theorem~\ref{thm_bound} cannot be substantially improved.

\begin{remark}\label{bottleneck}
We note that $\lambda_0, \lambda_1$ can be estimated efficiently. Indeed, let $e_{x_i} \in \RE^{\dvars \cup \Theta}$ be the $x_i$-th unit vector in $\mathbb{R}^{\dvars \cup \Theta}$, i.e., $e_{x_i}(x_j) = \delta_{i,j}$ where $\delta_{i,j}$ is the Kronecker delta. Then, if $y(t_0) = e_{x_i}$, Theorem~\ref{thm_linear} implies $y(t_1) = \Lambda(t_0,t_1) e_{x_i}$. Since $\Lambda(0, t_1) e_{x_i}$ is the $x_i$-column of $\Lambda(0,t_1)$ and $\Lambda(t_0,t_1) = \Lambda(0,t_1) \Lambda(0,t_0)^{-1}$, this shows that the matrices $\Lambda(t_0,t_1)$ can be computed by solving $|\dvars \cup \Theta|$ instances of the linear ODE system from Theorem~\ref{thm_linear} up to time $\hat{\tau}$.
\end{remark}
By calculating a bound $L > 0$ on $\max_{0 \leq t \leq \hat{\tau}} \lVert A(t) \rVert$ and by computing the matrices $\Lambda(t_i,t_j)$ for all time points $t_k$ underlying a fixed discretization step $\Delta t > 0$ of $[0;\hat{\tau}]$, the following can be shown.
\begin{lemma}\label{lem_lambda_num}
Together with $\lambda^+_0 = \max_i \lVert\Lambda(0,t_i)\rVert$ and $\lambda^+_1 = \max_{i \leq j} \lVert \Lambda(t_i,t_j) \rVert$, it holds that $\lambda_0 \leq \lambda_0^+ e^{L \Delta t}$ and $\lambda_1 \leq \lambda_1^+ [ 1 + L \Delta t ( e^{L \Delta t} + 1)]$.
\end{lemma}

The next result simplifies the constraints on $\delta$ from Theorem~\ref{thm_bound} if  $\deg(\hat{\calp}) \leq 3$.

\begin{lemma}\label{lem_muller}
In the case where $\deg(\hat{\calp}) \leq 3$, the constraint on $\delta$ of Theorem~\ref{thm_bound} simplifies to $\delta \leq \Big[ 2 \hat{\tau} \lambda_0 \lambda_1 \Big( d_2 + \sqrt{d_2^2 + \frac{2 d_3}{\lambda_1 \hat{\tau}}} \Big) \Big]^{-1}$.
\end{lemma}
The above lemma applies, for instance, to most biochemical systems, as discussed in Section~\ref{sec:ade}. The next result, instead, allows for an efficient estimation of $d_k$, with $2 \leq k \leq \deg(\hat{\calp})$.

\begin{lemma}\label{lem_eps0}
Given an extended PIVP $\hat{\calp}$ with variables $\dvars \cup \Theta$, let $\#_k(\hat{q}_i)$ be the number of degree $k$ monomials in $\nf(\hat{q}_i)$ and $D(\hat{q}_i,\hs)$ the largest coefficient of $\nf(\hat{q}_i)$ for configuration $\hs \in \RE^{\dvars \cup \Theta}$. With $C = \max_{0 \leq t \leq \hat{\tau}} \lVert x^{\hso}(t) \rVert$, $M = \max_{x_i \in \dvars} \max_{k \geq 2} \#_k(\hat{q}_i)$ and $D = \max_{x_i \in \dvars} D(\hat{q}_i,\hso)$, it suffices to set $d_k$ in Theorem~\ref{thm_bound} to $C^{\deg(\hat{\calp}) - k} M D$.
\end{lemma}
For the case of linear systems whose parameters are subject to perturbation, instead, the following lemma can be applied. It provides a sharper estimate on $d_2$ but comes at the price of more involved computation.
\begin{lemma}\label{lem_eps0_finer}
Given an extended PIVP $\hat{\calp}$ with variables $\dvars \cup \Theta$, the Hessian matrix $H^k = (H^k_{i,j})_{x_i,x_j}$ of $\hat{q}_k$ is given by $H^k_{i,j} = \partial_{x_i} \partial_{x_j} \hat{q}_k$. With this, $d_2$ can be chosen as $d_2 = \tfrac{1}{2} \cdot \max_{x_i \in \dvars \cup \Theta} \max_{0 \leq t \leq \hat{\tau}} \lVert H^i(x^{\hso}(t)) \rVert$.
\end{lemma}

\begin{example}\label{ex_mc4}
Since $\deg(\hat{\calp}) = 2$ in Example~\ref{ex_hatcalp}, coefficients $d_3,d_4,\ldots$ are zero and we only need to bound $d_2$. Moreover, the constraint in Theorem~\ref{thm_bound} simplifies to $\delta \leq (4 \hat{\tau} \lambda_0 \lambda_1 d_2)^{-1}$ thanks to Lemma~\ref{lem_muller}. By applying Lemma~\ref{lem_eps0}, instead, we see that it suffices to choose $d_2 = 2.00$ because $M = 2.00$ and $D = 1.00$. In the case of $\hat{\tau} = 3.00$, we thus get $\lambda_0 = \lambda_1 = 1.40$ which yields $\delta \leq 0.02$.
\end{example}

%% file: caseStudies.tex

\pdfoutput=1

\section{Evaluation}\label{sec:numerical}

\input{circuit}

\paragraph*{Discussion.}
In summary, the experimental results suggest that our bounding technique may complement the current state of the art in reachability analysis. Indeed, it has been shown to handle ODE systems of larger size,
but it provides a $\delta$ neighborhood that can explain perturbations up to ca 0.1\% at best. This makes our approach beneficial for the automatic  detection and abstraction from quasi-symmetries due to small uncertainties, e.g., measurement errors. On the other hand, algorithms such as those implemented in C2E2, CORA, and Flow$^\ast$ can theoretically cover arbitrarily larger initial uncertainties, but 
at a computational cost that blocked their applicability to our larger benchmarks. In future work it is worth investigating a possible combination of these techniques. 


%% file: circuit.tex
\pdfoutput=1



\begin{table}[t]
\begin{minipage}[b]{0.60\linewidth}
\centering
\includegraphics[width=60mm]{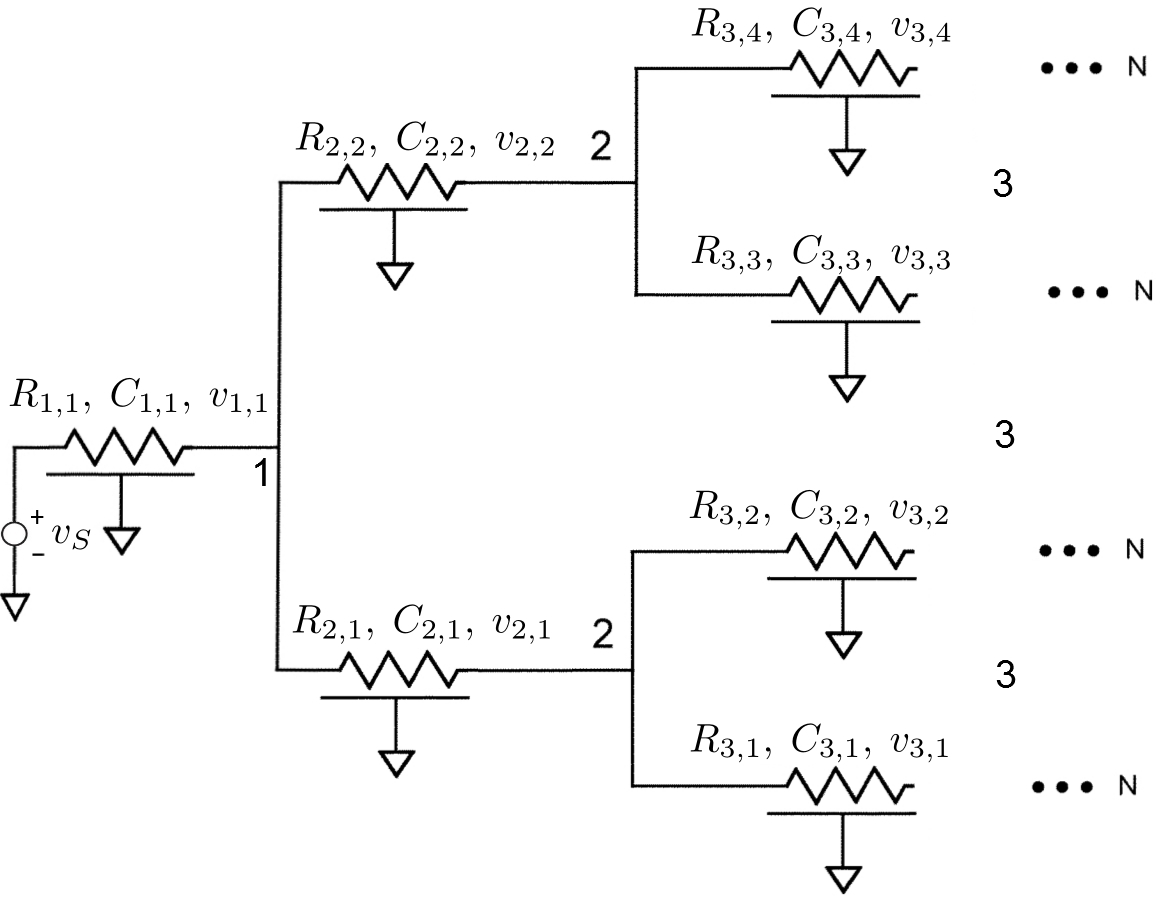}
\captionof{figure}{H-tree network adapted from~\cite{DBLP:journals/tvlsi/RosenfeldF07}.}
\label{fig1_htree}
\end{minipage}
\begin{minipage}[b]{0.40\linewidth}
\centering
    \begin{tabular}{rrrHH}
    \toprule
    \multicolumn{1}{c}{$i$} \ & $R^\ast_i$ (m$\Omega$) & \ \ $C^\ast_i$ \!\!\! (fF) & $\delta$ E--18 & $\lambda$ \\
    \midrule
    1 \ & 3.19    & \ 0.280 & N/A & N/A \\
    2 \ & 6.37    & \ 0.300 & \ 1.40E--3\ & 2.24 \\
    3 \ & 12.75   & \ 0.130 & \ 1.10E--3 \ & 2.59 \\
    4 \ & 25.50   & \ 0.140 & \ 5.21E--4 \ & 3.79 \\
    5 \ & 50.00   & \ 0.070 & \ 4.46E--4 \ & 4.01 \\
    6 \ & 100.00  & \ 0.070 & \ 4.46E--4 \ & 4.01 \\
    7 \ & 200.00  & \ 0.035 & \ 4.46E--4 \ & 4.01 \\
    8 \ & 400.00  & \ 0.035 & \ 4.46E--4 \ & 4.01 \\
    \bottomrule
    \end{tabular}
    \caption{Nominal parameters of electronic components at different depths $i$.}
    \label{table_components}
\end{minipage}\hfill
\end{table}

We consider a simplified (inductance free) version of a power distribution electrical network from~\cite{DBLP:journals/tvlsi/RosenfeldF07},  arranged as a tree called \emph{H-tree} (Figure~\ref{fig1_htree}). We let $N$ be the depth of the tree and denote the resistances and the capacitances at depth $i$ by $R_{i,k}$ and $C_{i,k}$, respectively. The source voltage is $v_s$, here assumed to be constant, $v_s = 2.0 \text{V}$. Then, the voltage across $C_{i,k}$, denoted by $v_{i,k}$, obeys the affine ODE
\begin{align}\label{eq_ex_ic_0}
\dot{v}_{1,1} & = \frac{v_S - v_{1,1}}{R_{1,1} C_{1,1}} - \frac{v_{1,1} - v_{2,1}}{R_{2,1} C_{1,1}} -\frac{v_{1,1} - v_{2,2}}{R_{2,2} C_{1,1}} , &
\dot{v}_{i,k} & = \frac{v_{i-1,l} - v_{i,k}}{R_{i,k} C_{i,k}} ,
\end{align}
where $1 \leq i \leq N$, $k = 1, \ldots, 2^{i - 1}$, and $l = \lceil k / 2 \rceil$, where $\lceil \cdot \rceil$ denotes the ceil function. Here we considered networks with depth up to $N = 8$. For depths $i \leq 4$, the nominal parameter values $R_i^\ast$ and $C_i^\ast$ were taken from~\cite{DBLP:journals/tvlsi/RosenfeldF07}; for $i \geq 5$, instead, we extrapolated them. The parameters are summarized in Table~\ref{table_components}.  

In~\cite{DBLP:journals/tvlsi/RosenfeldF07} the authors show that when the values of resistors and capacitors of any depth are equal, i.e., $R_{i,\cdot} \equiv R^\ast_i$ and $C_{i,\cdot} \equiv C^\ast_i$ then the network is \emph{symmetric}. That is, the voltages at the capacitors in any level are equal at all time points. Indeed, $\big\{ \{ v_{i,k} \mid 1 \leq k \leq 2^{i-1} \} \mid 1 \leq i \leq N \big\}$ is an exact BDE partition (with $N$ equivalence classes).

We now study the robustness of the symmetry under the realistic assumption that resistances and capacitances are only approximately equal. In particular, we test whether it is possible to explain quasi-symmetries when the parameters have tolerance  $\eta = 0.01\%$. This corresponds to a practical situation when components or measurements parameters enjoy high accuracy.
We considered networks of different size by varying the maximum depth $N$ from 2 to 8. For each size, in order to simulate a quasi-symmetric scenario we built 30 distinct ODE models by sampling values for $R_{i,k}$ and $C_{i,k}$ uniformly at random within $\eta$ percent from their nominal values. These repetitions  were made in order to avoid fixing a single instance that might unfairly favor our algorithm. To each model we  applied the $\varepsilon$-BDE reduction algorithm; choosing $\varepsilon = 6.00$E-4, it returned a quotient corresponding to a perfectly symmetrical case. The reduction times were below 0.5\,s in all cases. (Throughout this section, the runtimes reported were measured on a VirtualBox virtual machine running Ubuntu 64 bits over an actual 2.6 GHz Intel Core i5 machine with 4 GB of RAM.) Then, we computed the values of $\delta$ and $\lambda$ over a time horizon of 7 times units. This was chosen as a representative time point, for any $N$, of the transient state of the network (to account for the fact that, typically, circuits are analyzed in the time domain for transient analysis only).

The presence of uncertain parameters required us to transform the originally affine system~(\ref{eq_ex_ic_0}) into a polynomial system of degree two (by substituting each $1/(R_{i,k} C_{j,l})$ with a corresponding new state variable) with $2^{N + 1}$ states. This nonlinearity ruled out the application of standard over-approximation techniques for linear systems.

\input{tableCircut}
We present the results for the random model with the smallest value of $\delta$ in~Table~\ref{table_circuit_experiments}. The runtimes (second column) refer to the computation cost of the $\lambda$-$\delta$ pair. In all cases, $\delta$ turned out to be larger than the distance between the original model and its quotient $\lVert  \hs_0 - \hso \rVert_{} = \lVert  x^{\hs_0}(0) - x^{\hso}(0) \rVert_{}$ (shown in column $\lVert \cdot \rVert$). 
This demonstrates that the $0.01\%$ tolerance can be formally explained by approximate differential equivalence, as confirmed by the small values of the amplifiers $\lambda$.

We compared against C2E2, CORA, and Flow$^\ast$. (We did not compare our technique with other approaches applicable to nonlinear dynamics such as VNODE-LP~\cite{Nedialkov2011}, Ariadne~\cite{Benvenuti20088960} or HySAT/iSAT~\cite{Fraenzle07efficientsolving} because those have been compared to Flow$^\ast$ in the past~\cite{DBLP:conf/cav/ChenAS13}.) While for CORA and Flow$^\ast$ the comparison can be made directly, C2E2 seeks to decide whether a given set is reachable. In order to perform as fair a comparison as possible we chose unreachable sets generously far away from the over-approximations computed with our bound,  in order to ensure that C2E2 computes an over-approximation proving this.  In all three cases, the initial uncertain set was fixed so as to correspond to the ball around the initial condition of the reference model $\hso$ that included the original model $\hs_0$. This is the most favorable condition for the tools since it corresponds to the smallest uncertainty set which can provide a guaranteed error bound. Other settings of C2E2, CORA, and Flow$^\ast$ were chosen such to ensure successful termination. In the case of Flow$^\ast$ we used estimation remainder 0.10 and allowed for adaptive Taylor model degree between 2 and 6. In C2E2 we set the $K$ value to 2000. We used the time step 0.10 for C2E2 since this ensured safety for the models that could be analyzed. For CORA and  Flow$^\ast$ we set time step equal to 0.01 as this led to tight enough bounds. For our approach, instead, we used time step 0.023 because this ensured tight approximations of $\lambda_0$ and $\lambda_1$ via Lemma~\ref{lem_lambda_num}. In all cases the time-out was set to 3 hours. 

The comparison results are also reported in Table~\ref{table_circuit_experiments}. The over-approximations computed by C2E2 are not shown because they need not to be tight in order to verify the reachability problem. This is because C2E2 is refining an over-approximation only if this is necessary to decide safety. Thus, we report that C2E2 was able to verify a set to be safe, i.e., unreachable. For a network of depth $N$, the over-approximations for CORA and Flow$^\ast$ are reported as the maximal diameter of the flowpipe underlying $v_{N,1}$ across all time points. As such, it can be compared to the product $\lambda \cdot \lVert  \hs_0 - \hso \rVert_{}$ given by our bound, which is also explicitly reported in the table for the sake of easy comparability (column $\lambda \cdot \lVert  \cdot \rVert_{}$). Both CORA and Flow$^\ast$ reported tight bounds, of the same orders as ours. These correspond to at most ca. 1\% error on the observed variable, which cannot exceed the value of 2.0 (the source voltage applied to the network).  C2E2 did not terminate within the time-out with models with $N > 3$, while Flow$^\ast$ ran out of memory for $N > 4$; CORA, instead, failed to compute the symbolic Jacobian matrices for $N > 5$. Our approach, instead, timed out for $N = 8$. However, we wish to point out that our algorithm naturally applies to parallelization. Indeed, its bottleneck is in the computation of the set of linear ODE systems discussed in Remark~\ref{bottleneck}, which can be trivially solved independently from each other.

%% file: tableCircut.tex

\pdfoutput=1

\begin{table}[t]
\centering
\scalebox{0.93}{
    \begin{tabular}{cHH crrr rr rrrrrr}
    \toprule
\multicolumn{3}{c}{}  &  \multicolumn{5}{c}{\emph{Reference model}} &
     \multicolumn{6}{c}{\emph{Time (s)} \& \emph{Maximal Over-approximation}}
\\
\cmidrule(l){4-8} \cmidrule(l){9-14}
    \multicolumn{1}{c}{$N$} \ &
    &
    &
    \multicolumn{1}{c}{\ \emph{Time (s)}} & \multicolumn{1}{c}{$\lambda$}&  \multicolumn{1}{c}{$\delta$}   & \multicolumn{1}{c}{$\lVert  \cdot \rVert_{}$} & \multicolumn{1}{c}{$ \lambda \cdot \lVert  \cdot \rVert_{}$} 
    &
    \multicolumn{2}{c}{C2E2} & \multicolumn{2}{c}{CORA} & \multicolumn{2}{c}{FLOW$^\ast$}
     \\
    \midrule
    2 \ & 8    & 2  & 1.96E+0 & \  5.41 & \ 7.95E--4 & \ 4.43E--4 & 2.40E--3
    & \ 1.00E+1 & safe 
    & \ 4.82E+1 &  \ 8.50E--3 
    & \ 2.51E+1 & \ 8.21E--3
    \\
    3 \ & 16    & 3  & 3.51E+0   & \  6.27 & \ 6.33E--4 &\  7.42E--5 & \ 4.65E--4
    & \ 4.20E+2 & safe 
    & \ 1.42E+2 &  \ 7.80E--3 
    & \ 1.10E+2 & \ 9.74E--3
    \\
    4 \ & 32   & 4  & 7.75E+0 & \ 7.78 & \ 4.71E--4 & \ 2.17E--5  & \ 1.68E--4
    & \multicolumn{2}{c}{---} 
    &  \ 6.23E+2& \ 5.50E--3
    & \ 9.40E+2 & \ 1.16E--2
    \\
    5 \ & 64    & 5  & 2.46E+1 & \ 7.78 & \ 4.71E--4 & \ 3.31E--4  & \ 2.58E--3
    & \multicolumn{2}{c}{---} 
    & \ 5.39E+3 &\ 5.00E--3 
    & \multicolumn{2}{c}{---} 
        \\
    6 \ & 128   & 6  & 8.42E+1 &  \ 7.78 & \ 4.71E--4  & \ 8.97E--5 &  \ 6.98E--4
    & \multicolumn{2}{c}{---} 
    & \multicolumn{2}{c}{---} 
    & \multicolumn{2}{c}{---} 
        \\
    7 \ & 256   & 7   &  4.74E+2 & \ 7.78 & \ 4.71E--4 & \ 4.22E--4  & \ 3.28E--3
    & \multicolumn{2}{c}{---} 
    & \multicolumn{2}{c}{---}  
    & \multicolumn{2}{c}{---} 
    \\
    8 \ & 512   & 8  & \multicolumn{5}{c}{---}
    & \multicolumn{2}{c}{---} 
    & \multicolumn{2}{c}{---}  
    & \multicolumn{2}{c}{---} 
    \\
    \bottomrule
    \end{tabular}
    }
\caption{H-tree model results.}
    \label{table_circuit_experiments}
\end{table} 

%% file: ms.bbl
\begin{thebibliography}{10}
\providecommand{\url}[1]{\texttt{#1}}
\providecommand{\urlprefix}{URL }

\bibitem{DBLP:conf/cav/AbateBCK15}
Abate, A., Brim, L., Ceska, M., Kwiatkowska, M.Z.: {Adaptive Aggregation of
  Markov Chains: Quantitative Analysis of Chemical Reaction Networks}. In:
  {CAV}. pp. 195--213 (2015)

\bibitem{Althoff2013a}
Althoff, M.: {Reachability Analysis of Nonlinear Systems using Conservative
  Polynomialization and Non-Convex Sets}. In: {HSCC}. pp. 173--182 (2013)

\bibitem{Althoff2015a}
Althoff, M.: An introduction to {CORA} 2015. In: Proc. of the Workshop on
  Applied Verification for Continuous and Hybrid Systems (2015)

\bibitem{arand2012vivo}
Arand, J., Spieler, D., Karius, T., Branco, M.R., Meilinger, D., Meissner, A.,
  Jenuwein, T., Xu, G., Leonhardt, H., Wolf, V., et~al.: {In vivo control of
  CpG and non-CpG DNA methylation by DNA methyltransferases}. PLoS Genet  8(6)
  (2012)

\bibitem{Asarin2003}
Asarin, E., Dang, T., Girard, A.: {Reachability Analysis of Nonlinear Systems
  Using Conservative Approximation}. In: HSCC (2003)

\bibitem{Benvenuti20088960}
Benvenuti, L., Bresolin, D., Casagrande, A., Collins, P., Ferrari, A., Mazzi,
  E., Sangiovanni-Vincentelli, A., Villa, T.: {Reachability computation for
  hybrid systems with Ariadne}. \{IFAC\} Proceedings Volumes  41(2),  8960 --
  8965 (2008), {17th \{IFAC\} World Congress}

\bibitem{DBLP:conf/cav/BogomolovFGLPW12}
Bogomolov, S., Frehse, G., Grosu, R., Ladan, H., Podelski, A., Wehrle, M.: {A
  Box-Based Distance between Regions for Guiding the Reachability Analysis of
  SpaceEx}. In: {CAV}. pp. 479--494 (2012)

\bibitem{DBLP:conf/fossacs/Boreale17}
Boreale, M.: Algebra, coalgebra, and minimization in polynomial differential
  equations. In: {FOSSACS}. pp. 71--87 (2017)

\bibitem{worrell01:approximate}
van Breugel, F., Worrell, J.: Towards quantitative verification of
  probabilistic transition systems. In: ICALP (2001)

\bibitem{cttvPNAS}
Cardelli, L., Tribastone, M., Tschaikowski, M., Vandin, A.: {Maximal
  aggregation of polynomial dynamical systems}. Proceedings of the National
  Academy of Sciences  114(38),  10029 -- 10034 (2017)

\bibitem{popl16}
Cardelli, L., Tribastone, M., Tschaikowski, M., Vandin, A.: Symbolic
  computation of differential equivalences. In: POPL (2016)

\bibitem{DBLP:conf/cav/ChenAS13}
Chen, X., {\'{A}}brah{\'{a}}m, E., Sankaranarayanan, S.: {Flow*: An Analyzer
  for Non-linear Hybrid Systems}. In: {CAV}. pp. 258--263 (2013)

\bibitem{Danos200469}
Danos, V., Laneve, C.: Formal molecular biology. Theoretical Computer Science
  325(1),  69--110 (2004)

\bibitem{Donze2007}
Donz{\'e}, A., Maler, O.: {Systematic Simulation Using Sensitivity Analysis}.
  In: HSCC. pp. 174--189. Springer (2007)

\bibitem{Duggirala:2013:VAM:2555754.2555780}
Duggirala, P.S., Mitra, S., Viswanathan, M.: {Verification of Annotated Models
  from Executions}. In: {EMSOFT}. pp. 26:1--26:10. IEEE Press (2013)

\bibitem{E:2008aa}
E, W., Li, T., Vanden-Eijnden, E.: Optimal partition and effective dynamics of
  complex networks. PNAS  105(23),  7907--12 (Jun 2008)

\bibitem{DBLP:conf/cav/FanQM0D16}
Fan, C., Qi, B., Mitra, S., Viswanathan, M., Duggirala, P.S.: Automatic
  reachability analysis for nonlinear hybrid models with {C2E2}. In: {CAV}. pp.
  531--538 (2016)

\bibitem{1582235}
Girard, A., Pappas, G.: Approximate bisimulations for nonlinear dynamical
  systems. In: IEEE Conference on Decision and Control and European Control
  Conference (2005)

\bibitem{dsn13}
Iacobelli, G., Tribastone, M.: Lumpability of fluid models with heterogeneous
  agent types. In: DSN (2013)

\bibitem{DBLP:journals/tcs/IslamMBCFGSG15}
Islam, M.A., Murthy, A., Bartocci, E., Cherry, E., Fenton, F.H., Glimm, J.,
  Smolka, S.A., Grosu, R.: Model-order reduction of ion channel dynamics using
  approximate bisimulation. Theor. Comput. Sci.  599,  34--46 (2015)

\bibitem{IWASA01011989}
Iwasa, Y., Levin, S.A., Andreasen, V.: {Aggregation in Model Ecosystems II.
  Approximate Aggregation}. Mathematical Medicine and Biology  6(1),  1--23
  (1989)

\bibitem{KOZLOV1980223}
Kozlov, M., Tarasov, S., Khachiyan, L.: {The polynomial solvability of convex
  quadratic programming}. {USSR Computational Mathematics and Mathematical
  Physics}  20(5),  223 -- 228 (1980)

\bibitem{KuoWei:1969}
Kuo, J.C.W., Wei, J.: Lumping analysis in monomolecular reaction systems.
  {A}nalysis of approximately lumpable system. Industrial \& Engineering
  Chemistry Fundamentals  8(1),  124--133 (1969)

\bibitem{DBLP:conf/emsoft/LalP15}
Lal, R., Prabhakar, P.: Bounded error flowpipe computation of parameterized
  linear systems. In: {EMSOFT}. pp. 237--246 (2015)

\bibitem{Larsen19911}
Larsen, K.G., Skou, A.: Bisimulation through probabilistic testing. Inf.
  Comput.  94(1),  1--28 (1991)

\bibitem{LI1990977}
Li, G., Rabitz, H.: A general analysis of approximate lumping in chemical
  kinetics. Chemical Engineering Science  45(4),  977--1002 (1990)

\bibitem{DBLP:conf/cav/MajumdarZ12}
Majumdar, R., Zamani, M.: {Approximately Bisimilar Symbolic Models for Digital
  Control Systems}. In: {CAV}. pp. 362--377 (2012)

\bibitem{Fraenzle07efficientsolving}
{Martin Fr{\"a}nzle and Christian Herde and Tino Teige and Stefan Ratschan and
  Tobias Schubert}: {Efficient solving of large non-linear arithmetic
  constraint systems with complex boolean structure}. Journal on
  Satisfiability, Boolean Modeling and Computation  1,  209--236 (2007)

\bibitem{Pardalos1991}
Pardalos, P.M., Vavasis, S.A.: {Quadratic programming with one negative
  eigenvalue is NP-hard}. {Journal of Global Optimization}  1(1),  15--22
  (1991)

\bibitem{Pedersen:2010aa}
Pedersen, M., Plotkin, G.D.: A language for biochemical systems: Design and
  formal specification. In: IEEE/ACM TCBB XII, LNCS, vol. 5945, pp. 77--145.
  Springer (2010)

\bibitem{DBLP:journals/tvlsi/RosenfeldF07}
Rosenfeld, J., Friedman, E.G.: {Design Methodology for Global Resonant H-Tree
  Clock Distribution Networks}. {IEEE} Trans. {VLSI} Syst.  15(2),  135--148
  (2007)

\bibitem{Nedialkov2011}
S., N.: {Implementing a Rigorous ODE Solver Through Literate Programming}
  (2011)

\bibitem{DBLP:journals/tcs/TschaikowskiT14}
Tschaikowski, M., Tribastone, M.: {Tackling continuous state-space explosion in
  a Markovian process algebra}. Theor. Comput. Sci.  517,  1--33 (2014)

\bibitem{tac15}
Tschaikowski, M., Tribastone, M.: Approximate reduction of heterogeneous
  nonlinear models with differential hulls. IEEE TAC  (2016)

\end{thebibliography}
